\pdfoutput=1
\documentclass[%
 reprint,
 amsmath,amssymb,longbibliography,
 aps,
 prb,
]{revtex4-2}%
\usepackage{hyperref}
\usepackage{multirow}
\usepackage{booktabs}
\usepackage[dvipsnames]{xcolor}
\usepackage{dsfont}
\usepackage{nicefrac}
\usepackage{bm}
\usepackage{graphicx}
\usepackage{amsmath}
\usepackage{lipsum}
\usepackage{color}
\usepackage{wasysym}
\usepackage[normalem]{ulem}
\usepackage{float}

\usepackage{tikz}

\newcommand{\fR}{{\bf R}}
\newcommand{\fr}{{\bf r}}
\newcommand{\fk}{{\bf k}}

\newcommand{\fq}{{\bf q}}
\newcommand{\cd}{c^{\dagger}}

\graphicspath{{./figures/}}

\begin{document}
\title{Chern-insulator phases and spontaneous spin and valley order in a moir{\'e} lattice model for magic-angle twisted bilayer graphene} 

\author{Clara N. Brei\o}
\affiliation{Niels Bohr Institute, University of Copenhagen, 2100 Copenhagen, Denmark}

\author{Brian M. Andersen}
\affiliation{Niels Bohr Institute, University of Copenhagen, 2100 Copenhagen, Denmark}

\begin{abstract}
At a certain "magic" relative twist angle of two graphene sheets it remains a challenge to obtain a detailed description of the proliferation of correlated topological electronic phases and their filling-dependence. We perform a self-consistent real-space Hartree-Fock study of an effective moir{\'e} lattice model to map out the preferred ordered phases as a function of Coulomb interaction strength and moir{\'e} flat-band filling factor. It is found that a quantum valley Hall phase, previously discovered at charge neutrality, is present at all integer fillings for sufficiently large interactions. However, except from charge neutrality additional spontaneous spin/valley polarization is present in the ground state at nonzero integer fillings, leading to Chern-insulator phases and anomalous quantum Hall effects at odd filling factors, thus constituting an example of interaction-driven nontrivial topology. At weaker interactions, all nonzero integer fillings feature metallic inhomogeneous spin/valley ordered phases which may also break  additional point group symmetries of the system. We discuss these findings in the light of previous theoretical studies and recent experimental developments of magic-angle twisted bilayer graphene.   
\end{abstract}

\date{\today}
\maketitle

\section{Introduction}
Magic-angle twisted bilayer graphene (MATBG) provides a tunable platform for studying the properties of strongly-correlated electrons on moir{\'e} superlattices~\cite{Cao2018_Mott,Cao2018_SC,Andrei2020,Balents2020}. The moir{\'e} lattice is generated by the relative twist which may, at certain angles, lead to very narrow low-energy minibands hosting electrons with large interactions relative to their respective kinetic energy~\cite{Santos2007,Morell2010,Bistritzer2011}. At the magic angle of $\theta \simeq 1.1^\circ$ the flat bands are well separated from the dispersive bands at higher energies. An incipient Berry curvature of the moir{\'e} flat bands additionally imprints the correlated electronic states with nontrivial topology. The resulting emergent electronic behavior includes both unusual normal state properties and the formation of different electronic phases exhibiting magnetic and superconducting signatures at low temperatures~\cite{Andrei2020,Balents2020}. Thus MATBG is interesting in its own right, and may also help shed light on the more general outstanding questions pertaining to strange-metal behavior and the origin of unconventional superconductivity~\cite{Cao2018_Mott,Cao2018_SC}. Therefore it is important to understand this system and elucidate the nature of the emergent electronic phases appearing in MATBG moir{\'e} superlattices.  

Experimentally, a series of mainly transport and scanning tunneling spectroscopy (STS) measurements on MATBG have investigated its phase diagram as a function of temperature and electronic density, reporting correlated insulating states at integer filling factors separated by superconducting domes at the lowest temperatures~\cite{Cao2018_Mott,Cao2018_SC,Lu2019,Yankowitz2019,Xie2019,Kerelsky2019,Jiang2019,Stepanov2020,Saito2020,Park2021}. At some filling factors of the moir{\'e} flat bands, there is evidence for magnetism and quantum anomalous Hall effect~\cite{Lu2019,Sharpe2019,Serlin2020,Stepanov2020,Stepanov2021,Grover2022}. Other correlated states under consideration include Chern insulators and nematic phases also at integer moir{\'e} band filling~\cite{Nuckolls2020,Wu2021,Pierce2021,Das2021}, and translation symmetry-breaking spin- or charge-density wave order at certain half-integer fillings~\cite{Bhowmik2022}. Local electronic compressibility and STS measurements recently revealed a sequence of distinct phase transitions near the integer fillings of the moir{\'e} unit cell, establishing a cascade of transitions associated with split-off bands of particular spin or valley character~\cite{Zondiner2020,Wong2020,Choi2021}. The detailed properties of the phase diagram appears to depend rather sensitively on the twist angle, substrate potential, or the degree of alignment of one of the graphene sheets with the hexagonal boron nitride encapsulation layers, which breaks an inherent $C_2$ symmetry of the graphene bilayer itself~\cite{Yankowitz2019,Lu2019,Serlin2020,Stepanov2021}. The twist-generated correlated phases have also been explored in other graphene-based systems, including e.g. twisted trilayers or twisted double bilayer graphene. Such stacked graphene sheets also feature rich phase diagrams with superconductivity and tunable insulator states, sensitive to both twist angle and electric displacement field~\cite{Chen2019,Chen2019_N,Cao2020_TDBG,Shen2020_TDBG,Liu2020_TDBG}.

Theoretically, the existence of insulating states in MATBG has been addressed within a large variety of techniques and approximations~\cite{PoPRX2018,Isobe_2018,Thomson2018,Dodaro2018,Ochi2018,KangVafek2019,Seo2019,LiaoPRL2019,Zhang_PRB2019,Wolf2019,Song_2019,Liu_PRX_2019,Sboychakov2019,Classen2019,ZhangPRR2019,Bultinck_PRX2020,Chichinadze2020,Bultinck_PRL2020,Kang_PRB2020,Rademaker_2020,Cea_2020,Xie_2020,Zhang_Zi2020,Repellin_2020,Liu2021,Liu_PRR2021,Lian2021,ChenMeng2021,Xie2021,Shavit_2021,KangPRL2021,Potasz2021,ClaraPRX2021,Klebl2021,Kwan_2021,Chichinadze2022,Wagner2022,Song_2022}. Obtaining a detailed quantitative theoretical description of the origin and interplay of different flavor-ordered phases and their dependence on e.g. filling and twist angle is still ongoing. These efforts are important not only from the perspective of understanding transport and tunneling measurements, but potentially also for identifying the "normal state" out of which superconductivity emerges at lower temperatures. From the perspective of Hartree-Fock (HF) approximated interactions, such an approach can lead to insulating behavior from induced order causing a spectral gap, typically arising from one or more spontaneously broken symmetries of the original Hamiltonian. Application of HF theory to handle electronic interactions has been widely applied to MATBG in order to describe the correlated insulator phases~\cite{Xie_2020,Bultinck_PRX2020,Xie2021,Liu2021,Shavit_2021,ChenMeng2021,ClaraPRX2021,Liu_PRR2021,Kwan_2021,Wagner2022}. Such studies have proposed, depending on the band filling factor, a rather large variety of spontaneously symmetry-broken candidate phases relevant for MATBG. These include spin- and valley-ferromagnetic phases, intervalley coherent insulators, valence-bond solids, nematic semi-metals, quantum valley Hall and quantum spin Hall insulators, incommensurate Kekul{\'e} spiral order, and various other forms of inhomogeneous density-wave order~\cite{PoPRX2018,Isobe_2018,Thomson2018,Dodaro2018,Ochi2018,KangVafek2019,Seo2019,LiaoPRL2019,Zhang_PRB2019,Wolf2019,Song_2019,Liu_PRX_2019,Sboychakov2019,Classen2019,ZhangPRR2019,Bultinck_PRX2020,Chichinadze2020,Bultinck_PRL2020,Kang_PRB2020,Rademaker_2020,Cea_2020,Xie_2020,Zhang_Zi2020,Repellin_2020,Liu2021,Liu_PRR2021,Lian2021,ChenMeng2021,Xie2021,Shavit_2021,KangPRL2021,Potasz2021,ClaraPRX2021,Klebl2021,Kwan_2021,Chichinadze2022,Wagner2022}.

Most theoretical works depart from the Bistritzer-MacDonald model with extended continuum wave functions and Coulomb interactions projected into the continuum states, including a number of the remote bands. A complementary Wannier description to study the correlated phases of MATBG was suggested by Kang and Vafek where the Coulomb interaction is projected onto the localized Wannier states of the four spin-degenerate moir{\'e} flat bands~\cite{KangVafek2019}. Due to Wannier obstruction, symmetry requirements within this framework lead to crucial non-local assisted-hopping-like interaction terms in real space~\cite{KangVafek2019}. A previous work studied the role of these assisted-hopping interactions on the ground state of MATBG at charge neutrality via HF mean-field studies and quantum Monte Carlo (QMC) simulations~\cite{ClaraPRX2021}. It was found that the normal state Dirac semi-metal phase was unstable and insulating already at weak coupling, which was a consequence of an induced quantum valley Hall (QVH) phase~\cite{ClaraPRX2021}. Upon variation of the relative strength of the assisted-hopping interactions and overall interaction strength, several other insulating phases were found to emerge. These include 1) on-site inter-valley coherence (IVC) order, which breaks the spin-valley SU(4) symmetry of the interacting part of the model, and 2) an insulating columnar valence bond solid state. A main result of Ref.~\onlinecite{ClaraPRX2021} was the importance of the assisted-hopping term in stabilizing the QVH and IVC states. It was concluded therefore, that an experimental observation of these quantum states at charge neutrality would provide evidence for the importance of non-local topologically driven interactions in MATBG.

Here, we extend the HF studies of Ref.~\onlinecite{ClaraPRX2021} to address other filling factors of the moir{\'e} flat bands. We perform an unrestricted mean-field real-space study which allows for both homogeneous and inhomogeneous solutions. In the former case, we have benchmarked the results against a momentum-space formalism, which additionally allows for determination of the interaction-renormalized band structure. Our theoretical study focuses on the case of zero external magnetic field. At charge neutrality we find that interactions induce a gapped homogeneous QVH phase in agreement with earlier studies of the same model~\cite{ClaraPRX2021}. At other integer fillings, the QVH phases coexists with spin/valley-flavor symmetry-broken order, consisting of spin/valley polarized phases for interaction strengths larger than approximately the bandwidth. We analyze the topological properties of the resulting gapped phases where the spontaneous order has lifted the degeneracy of the flavor symmetries of the bare band. For weaker interaction strengths at nonzero integer filling factors, or at certain half-integer fillings, the model prefers metallic inhomogeneous spin- and valley-ordered phases. We analyse these results in light of recent experimental developments of MATBG.

\begin{figure}[t]
\centering
\includegraphics[angle=0,width=1.0\linewidth]{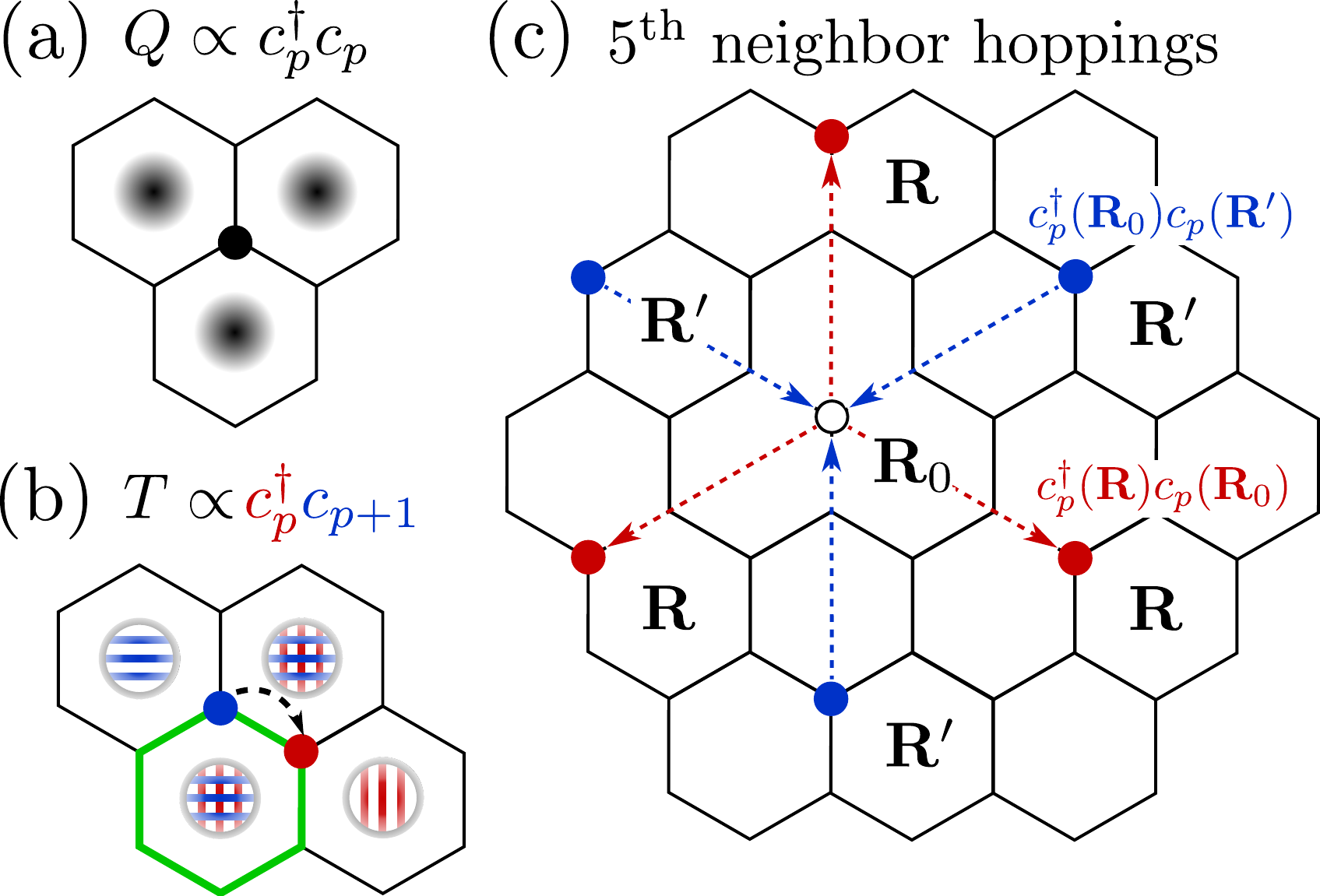}
\caption{(a) Illustration of the cluster charge term $Q(\fR)$ entering Eq.~(\ref{eq:Hint}). Shaded gray areas is the density distribution of a Wannier orbital at the honeycomb site marked by a black dot. (b) Schematic of the assisted-hopping term $T(\fR)$. Blue horizontal (red vertical) lines show the density distribution of a Wannier orbital centered at the blue (red) honeycomb site. The overlap within the hexagon highlighted in green is finite and non-negligible. (c) Fifth NN hoppings in $H_0$. Hoppings marked in blue are the terms in the $\fR'$-summation for $\fR = \fR_0$ and hoppings marked in red are included when the $\fR$-summation reaches hexagons marked by $\fR$ and $\fR'=\fR_0$.}
\label{fig:1}
\end{figure}

\section{Model and method}

Constructing an interacting lattice Hamiltonian modelling MATBG is complicated due to the fragile topology of the narrow bands preventing a faithful representation in terms of standard localized Wannier orbitals, a property known as Wannier obstruction~\cite{Po2018,Po2019}. To overcome this obstruction one must either add (trivial) remote bands by including additional orbitals or implement some of the required symmetries non-locally by including longer-range interactions. The model presented by Kang and Vafek~\cite{KangVafek2019} takes the latter approach and implements the $C_2 \mathcal{T}$-symmetry non-locally, where $C_2$ refers to a twofold rotation with respect to the $z$ axis and $\mathcal{T}$ denotes time reversal. This procedure allows for a projection of the screened Coulomb potential to the appropriate low-energy Wannier orbitals residing on a bipartite honeycomb lattice~\cite{Kang_2018,KangVafek2019}. The projection was performed in Ref.~\onlinecite{KangVafek2019} and led to the following interaction term
\begin{align}
    H_{\rm int} = \frac{U}{2} \sum_{\fR} \left(Q(\fR) + T(\fR)\right)^2, \label{eq:Hint}
\end{align}
where
\begin{equation}
Q(\fR) = \frac{1}{3} \sum^5_{p = 0} \sum_{\tau,\sigma} \cd_{p \tau \sigma}(\fR) c_{p \tau \sigma}(\fR),
\end{equation}
and 
\begin{equation}
T(\fR) = \alpha \sum^5_{p = 0} \sum_{\tau,\sigma} (-1)^{p} \left(\cd_{p \tau \sigma}(\fR)c_{p+1 \tau \sigma}(\fR) + \rm{H.c.} \right). \label{eq:AssHop}
\end{equation}
Here, $U$ sets the overall interaction strength, $\fR$ labels the position of a hexagon and $p$ labels the six sites in each hexagon, see Fig.~\ref{fig:1}. The factor of 1/3 in $Q(\fR)$ accounts for the triple counting and $\tau = \pm 1$ ($\sigma = \uparrow,\downarrow$) is the valley (spin) degree of freedom of the Wannier orbitals represented by the operators $c_{p\tau\sigma}$. As illustrated in Fig.~\ref{fig:1}(a,b) the low-energy Wannier orbitals in MATBG exhibit an inherent non-locality where, instead of the usual exponentially localized, single-peaked, wave function, an orbital centered at site $p$ exhibits three symmetric peaks in the adjacent hexagon centers~\cite{Koshino2018,Kang_2018}. 
In Fig.~\ref{fig:1}(a) we depict the cluster charge interaction term $Q(\fR)$, a usual density term apart from the non-local wave function distribution. Thus the contribution to $H_{\rm int}$ from $Q^2(\fR)$ is analogous to a hexagon-centered density-density Hubbard term containing onsite and up to third nearest-neighbor interactions in the projected model orbitals.

The assisted-hopping interaction term $T(\fR)$ is schematically shown in Fig.~\ref{fig:1}(b), where horizontal blue (vertical red) lines indicate the density distribution of a Wannier orbital residing at the blue (red) honeycomb site. While the total overlap of the two neighboring orbitals must be zero from orthogonality requirements, the overlap within a single hexagon (highlighted in green in Fig.~\ref{fig:1}(b)) can be finite. This term is topological in nature and arises from the (symmetry required) non-locality of the Wannier states~\cite{KangVafek2019}. In Ref.~\onlinecite{KangVafek2019} the authors find that the overlap integral $\alpha \sim 1/3$, yielding comparable interaction strengths for the $Q(\fR)$ and $T(\fR)$ terms. The assisted-hopping interactions are pivotal to all results presented in this paper.

The kinetic terms of the Hamiltonian were proposed e.g. in Refs.~\onlinecite{YuanFu2018,Koshino2018,Kang_2018}. The effective tight-binding model is defined on the AB/BA honeycomb lattice of the moir{\'e} superlattice and includes nearest-neighbor (NN) as well as complex fifth NN hopping. The model has eight narrow bands reflecting the spin, valley and sublattice degrees of freedom. In the notation of Eq.~\eqref{eq:Hint}, the minimal tight-binding kinetic part reads
\begin{widetext}
\begin{align}
    H_0 &= \sum_{\fR}\sum_{p,\tau,\sigma} \Big[ -\frac{\mu}{3}\cd_{p \tau \sigma}(\fR) c_{p \tau \sigma}(\fR) + t_1 e^{(-1)^{p-1} i \tau \phi}\cd_{p \tau \sigma}(\fR) c_{p+1 \tau \sigma}(\fR) +\frac{1}{3} \sum_{\fR'} \left( (t_2 -\tau i t'_2) \cd_{p \tau \sigma}(\fR) c_{p \tau \sigma}(\fR') + \rm{H.c.} \right) \Big],
\end{align}
\end{widetext}
where $\mu$ is the chemical potential, $t_1$ is the NN hopping amplitude and the phase factor, $e^{(-1)^{p-1}i\tau\phi}$, arises from a convenient gauge transformation ensuring $\alpha \in \mathbb{R}$ in Eq.~\eqref{eq:AssHop}~\cite{KangVafek2019}. In the last term $\fR'$ refers to three $C_3$ related NNN hexagons to $\fR$. As the site indices, $p$, within $\fR$ and $\fR'$ are identical, these terms are fifth NN hoppings with real ($t_2$) and imaginary ($t'_2$) hopping amplitudes, respectively, see Fig~\ref{fig:1}(c).

The validity of mean-field approaches in describing interacting electrons in moir{\'e} flat bands is questionable. However, as mentioned above, HF theory has been rather widely applied and provided insight into the possible ordered phases. Additionally, relevant for the present model, HF results were shown to agree surprisingly well with QMC simulations at charge neutrality. Motivated by these previous results, we proceed by performing an unrestricted HF decoupling of Eq.~\eqref{eq:Hint} yielding
\begin{widetext}
\begin{align}
    H^{\rm HF}_{\rm int} = U \sum_{\fR} \left(\bar{n}(\fR)(Q(\fR) + T(\fR)) - \sum_{\rm all} \left[\sum_{n,m} \alpha_n(p') \alpha_m(p)\langle \cd_{p'+n\tau'\sigma'} c_{p+m\tau\sigma} \rangle \right] \cd_{p\tau\sigma}c_{p'\tau'\sigma'} \right), \label{eq:HintHF}
\end{align}
with
\begin{align}
    \bar{n}(\fR) = \sum_{p',\tau',\sigma'}\left(\frac{1}{3}\langle\cd_{p'\tau'\sigma'}c_{p'\tau'\sigma'}\rangle+\alpha(-1)^{p'}[\langle\cd_{p'\tau'\sigma'}c_{p'+1\tau'\sigma'}\rangle+\langle\cd_{p'+1\tau'\sigma'}c_{p'\tau'\sigma'}\rangle]\right).
\end{align}
\end{widetext}
Here, $\sum_{\rm all} = \sum_{p,p'}\sum_{\tau,\tau'}\sum_{\sigma,\sigma'}$ and $n,m\in\{-1,0,1\}$. We have  suppressed the $\fR$ dependence of the operators for clarity and defined $\pmb{\alpha}(p) \equiv \left(\alpha_{-1}(p),\,\alpha_0(p),\,\alpha_1(p)\right) = \left(\alpha(-1)^{p-1},\,1/3,\,\alpha(-1)^p\right)$. Note from the definition of $\pmb{\alpha}(p)$ that setting $\alpha \neq 0$ inevitably introduces non-locality in the Fock exchange terms since e.g. NNN hopping, $\cd_{p}c_{p' = p+2}$, will contain contributions from onsite mean fields, $\langle \cd_{p+2-1} c_{p+1} \rangle$ for $n = -1$ and $m = 1$. 

The unrestricted HF decoupling leading to Eq.~\eqref{eq:HintHF} allows for spontaneous breaking of all symmetries including translational invariance, yielding a total of $R_{\rm tot}/3 \times 24 \times 24 \sim \mathcal{O}(10^5)$ different mean fields to be solved for typical systems sizes of $R_{\rm tot} = 300$.
The resulting decoupled Hamiltonian, $H = H_0 + H^{\rm HF}_{\rm int}$ is solved self-consistently with unbiased initial conditions. For the results reported on this paper, we have defined the convergence criterion to be $\sum^N_{n=1}|E_n(m-1)-E_n(m)|<N\cdot10^{-10}t_1$, where $m$ denotes an iteration counter and $N = R_{\rm tot}/3 \times 24 \times 24$ is the total number of eigenenergies $E_n$. Guided by Refs.~\onlinecite{YuanFu2018, KangVafek2019} we fix $\{t_1,\,t_2,\,t'_2,\,\phi,\,\alpha\}~=~\{1.0,\,0.025,\,0.1,\,0.743\pi,\,0.23\}$ while the filling is controlled by varying $\mu$ during the iterations until the desired filling factor is reached. As usual, we denote the filling factor by $\nu$ which is related to the carrier density $n$ by $\nu=4 n/n_s$, where $n_s$ is the density of the filled moir{\'e} superlattice flat bands. The temperature is set to $T = 2.5\times10^{-5}t_1$ in all computations below. 

\section{Results}

\subsection{Phase diagram: homogeneous phases}\label{sec:QVH}
\begin{figure}[t]
\centering
\includegraphics[angle=0,width=0.99\linewidth]{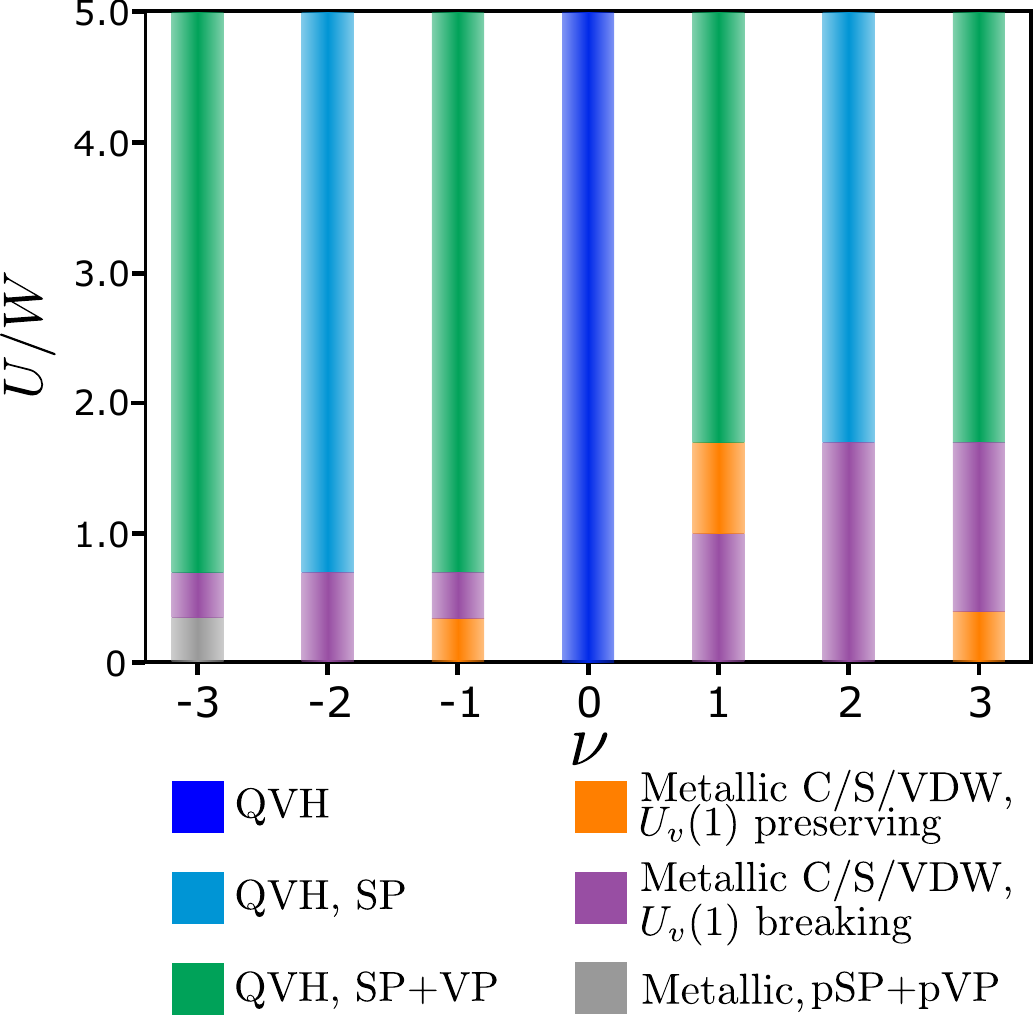}
\caption{Phase diagram indicating the preferred order at integer filling factors as a function of interaction strength $U/W$, where $W = 6t_1$ is the bare bandwidth. Charge neutrality features only a QVH phase in this range of interactions. The other integer fillings exhibit different ferromagnetic order with spin/valley polarization beyond a certain interaction threshold. Below this critical interaction value, all nonzero integer fillings prefer different inhomogeneous metallic phases. Note that the metallic phase marked in grey at $\nu = -3$ is only partially spin and valley polarized (pSP, pVP).} \label{fig:2}
\end{figure}

\begin{figure*}[t]
\centering
\includegraphics[angle=0,width=0.99\linewidth]{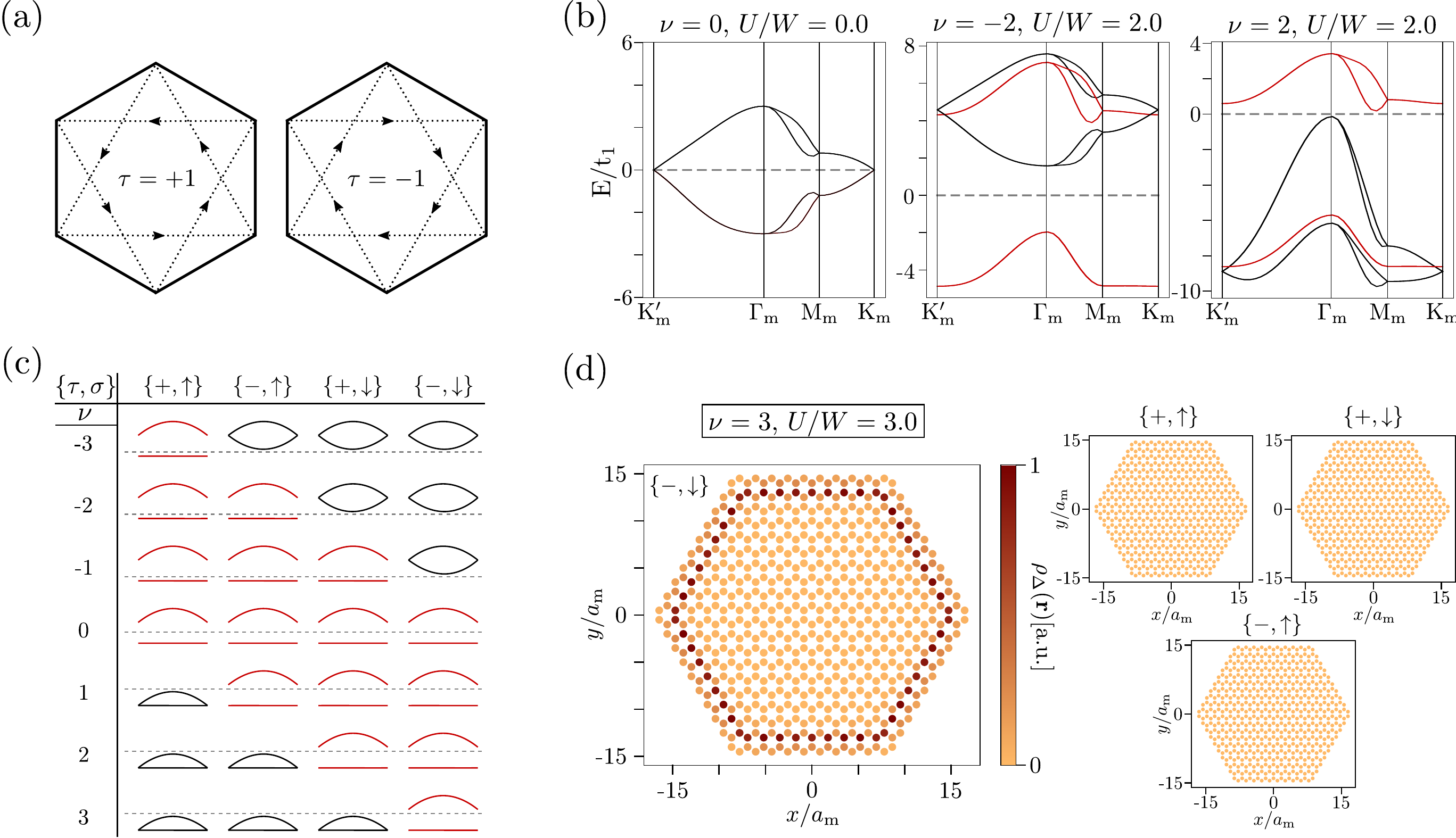}
\caption{(a) Illustration of emergent imaginary NNN hoppings in the QVH phases. The arrow direction indicate the direction of the corresponding loop currents dictated by the valley flavor. (b) Bare bands at charge neutrality (left panel) and renormalized bands at interaction strength $U/W = 2.0$ for $\nu = -2$ (center panel) and $\nu = +2$ (right panel). Gapless (Gapped) band sectors are marked in black (red). (c) Schematic of successive filling of bands and resulting gap opening/closing at all integer filling factors of the four decoupled Haldane sectors. Each sector is characterized by the spin and valley flavor $\{\tau,\sigma\}$. (d) In-gap local density of states, $\rho_{\Delta}(\fr)$, of the spin- and valley-polarized QVH phase at $\nu = 3.0$ and $U/W = 3.0$. A single topologically protected edge state is present in the only remaining gapped sector $\{-,\downarrow\}$. The results in (d) are computed by opening the boundaries in a result converged with periodic boundary conditions.} \label{fig:3}
\end{figure*}

We start by presenting a main result of the paper in Fig.~\ref{fig:2}: the full phase diagram as a function of interaction strength at all integer fillings. Focusing first on homogeneous phases found at sufficiently large interaction strengths $U/W$, the emergent orders are: 
\begin{itemize}
    \item Quantum valley Hall (QVH) order consisting of imaginary next-nearest-neighbor (iNNN) hopping, $\mathrm{Im}\langle \cd_{p\tau\sigma} c_{p+2\tau\sigma}\rangle \neq 0$, see Fig.~\ref{fig:3}(a).
    \item Fully spin-polarized (SP) order characterized by ferromagnetic onsite mean fields, $\langle n_{\uparrow}\rangle - \langle n_{\downarrow}\rangle = 1,2$ with $\langle n_{\sigma} \rangle = \frac{1}{3}\sum_{p\tau}\langle \cd_{p\tau\sigma}c_{p\tau\sigma} \rangle$, i.e. an excess of either one or two spin-up electrons per moir{\'e} unit cell.
    \item Fully valley-polarized (VP) order defined by $\langle n_{+1} \rangle - \langle n_{-1} \rangle = \pm1$ with $\langle n_{\pm1} \rangle =  \frac{1}{3}\sum_{p\sigma} \langle \cd_{p,\pm1,\sigma}c_{p,\pm1,\sigma} \rangle$.
    \end{itemize}
Despite the rich variety of possible symmetry breaking in MATBG, we find QVH order to be present in the ground state across all integer fillings in a wide range of interaction strengths. In Fig.~\ref{fig:2} the QVH phase is indicated by dark blue (non-symmetry-breaking), light blue when coexisting the spin polarization, and green when coexisting with both spin and valley polarization. At charge neutrality, the QVH phase is induced even for very small interaction strengths in agreement with Ref.~\onlinecite{ClaraPRX2021}, whereas at all nonzero integer fillings stabilization of the QVH phase requires a critical value of $U/W$.  The emergent iNNN hopping leads to an insulating gap, analogous to the Haldane model~\cite{Haldane1988}. The QVH phase preserves the $U_v(1)$ symmetry yielding a total of four decoupled sectors (valley$\times$spin) of the Haldane bands. The direction of the iNNN hopping is dictated by the valley flavor alone, see Fig.~\ref{fig:3}(a). Finally, we note that both $H_0$ and $H_{\rm{int}}$ break particle-hole symmetry resulting in a significant particle-hole asymmetry of the phase diagram in Fig.~\ref{fig:2}~\cite{ClaraPRX2021}. 

\subsection{Resulting band structure}\label{sec:QVHbands}
We turn now to the band structure associated with the homogeneous phases present in the phase diagram of Fig.~\ref{fig:2}. The leftmost panel in Fig.~\ref{fig:3}(b) shows the bare bands, exhibiting the well-known semi-metallic Dirac dispersion~\cite{YuanFu2018,Koshino2018,Kang_2018}. The center and right panels of Fig.~\ref{fig:3}(b) display the band structure at $U/W = 2.0$ and $\nu = -2,2$, respectively. From the renormalized bands at $\nu = -2$ it is evident that the interactions shift the minima of the occupied bands from $\Gamma_{\rm m}$ to $K_{\rm m}$ ($K'_{\rm m}$). The shift is caused by the assisted-hopping interactions which, as mentioned in the previous section, introduces longer-range hoppings leading to a non-trivial structure of $H_{\rm int}$. Combining this interaction-originated structure with the bare bands leads to distinct renormalization of the four lower and upper bands, respectively. At weak interactions ($U/W \lesssim 1.0$) the renormalization of the four lower bands suppresses the bandwidth significantly followed by the inversion evident in the center panel of Fig.~\ref{fig:3}(b) at larger interaction strength. On the contrary, since the upper four bands exhibit a similar momentum dependence as $H_{\rm int}$, the interactions will enhance the bandwidth of the these bands for all interaction strengths leading to reduced gap sizes as well as larger critical $U/W$ for all $\nu > 0$, see right panel of Fig.~\ref{fig:3}(b). 

By further inspection of Fig.~\ref{fig:3}(b) several similarities between the band structures at $\nu = -2$ and $2$ can be identified. Both structures exhibit significant energy gaps in two out of four Haldane sectors (red sectors in Fig.~\ref{fig:3}(b)) in agreement with the respective filling. However, more strikingly, we note that these two sectors are identical at the two different fillings. As such, the sole qualitative discrepancy between $\nu = -2$ and $2$ is whether the remaining two gapless sectors (black sectors in Fig.~\ref{fig:3}(b)) are empty ($\nu = -2$) or full ($\nu = 2$). The resemblance between the QVH band structures in Fig.~\ref{fig:3}(b) serve as an example of the general interpretation of the QVH phases across all integer fillings as illustrated in Fig.~\ref{fig:3}(c). At $\nu = -4$ the system consists of four empty, gapless and decoupled band sectors characterized by the valley and spin degrees of freedom, $\{\tau,\sigma\}$. Doping the narrow bands to the first commensurate filling ($\nu = -3$) results in an accumulation of all electrons into a single flavor. This accumulation is energetically favored as it allows for a gap opening in the corresponding band sector at the expense of spontaneously generated loop currents (i.e. iNNN hoppings).
Upon further electron filling, the mechanism repeats at each integer filling where a gap is introduced in one of the (previously gapless) sectors, until all four flavors are fully gapped at charge neutrality ($\nu = 0$). Increasing the electron density further, the mechanism reverses such that at each integer filling, an additional conduction band is fully occupied. Since all anomalous hopping terms (i.e. spin/valley flipping hoppings) vanish, the electrons in the fully occupied sectors are effectively  blocked, leading to a gap closing as iNNN hopping is prohibited. We attribute this behavior to be related to the cascade transitions seen experimentally~\cite{Zondiner2020,Wong2020,Choi2021}.

\subsection{Topology and edge states of the insulating phases}

The emergence of loop currents and spin polarization naturally leads to a consideration of  time-reversal symmetry breaking (TRSB). As all four flavor sectors remain decoupled in the QVH phases throughout the entire doping range, it is evident that the QVH order is accompanied by flavor polarization at all nonzero integer fillings. Specifically, all even filling factors ($\nu = \pm 2$) are spin polarized while all odd filling factors ($\nu = \pm 1, \pm 3$) exhibit both valley and spin polarization. Thus, time-reversal symmetry is spontaneously broken in the QVH phases at all nonzero integer filling factors. Since TRSB is the crucial element to obtain nontrivial topological bands in the Haldane model, it is reasonable to assume that the QVH phases discussed here also exhibit nontrivial topology. To verify this explicitly, we compute the Chern numbers at all integer fillings by evaluating the usual momentum integral 
\begin{align}
    C_{n} = \frac{1}{2\pi} \int_{mBZ}\!d^2\fk \, {\rm Tr}[\Omega^{n}(\fk)], 
\end{align}
where the Berry curvature is defined by
\begin{align}
    \Omega^{n} (\fk) = i P_{n}(\fk) \left[\partial_{k_x} P_{n}(\fk), \partial_{k_y} P_{n} (\fk)\right],
\end{align}
and $P_{n}$ is the projector to the corresponding band defined by the flavor sector and particle/hole character. The resulting Chern numbers are $C_n = \pm1$ ($\mp1$) for valence (conduction) band in $\tau=\pm1$ insulating sectors. The invariants are independent of spin direction in agreement with the purely valley-defined hopping direction depicted in Fig.~\ref{fig:3}(a). Not surprisingly, the non-gapped, fully occupied sectors yield $C_n = 0$ for both bands verified by introducing a perturbative splitting prior to the Chern number computation. Thus, all half-filled sectors host a single edge state while fully occupied sectors do not. An example of this property is shown in Fig.~\ref{fig:3}(d) which displays the in-gap local density of states, $\rho_{\Delta}(\fr)$, for all four sectors in the QVH phase at $\nu = 3$. As the propagation direction of the edge states are set by the valley degree of freedom, these findings yield quantum anomalous Hall phases, i.e. nonzero total Chern number, for all valley polarized QVH phases, that is, for all odd-integer fillings.

\begin{figure}[t]
\centering
\includegraphics[angle=0,width=1.0\linewidth]{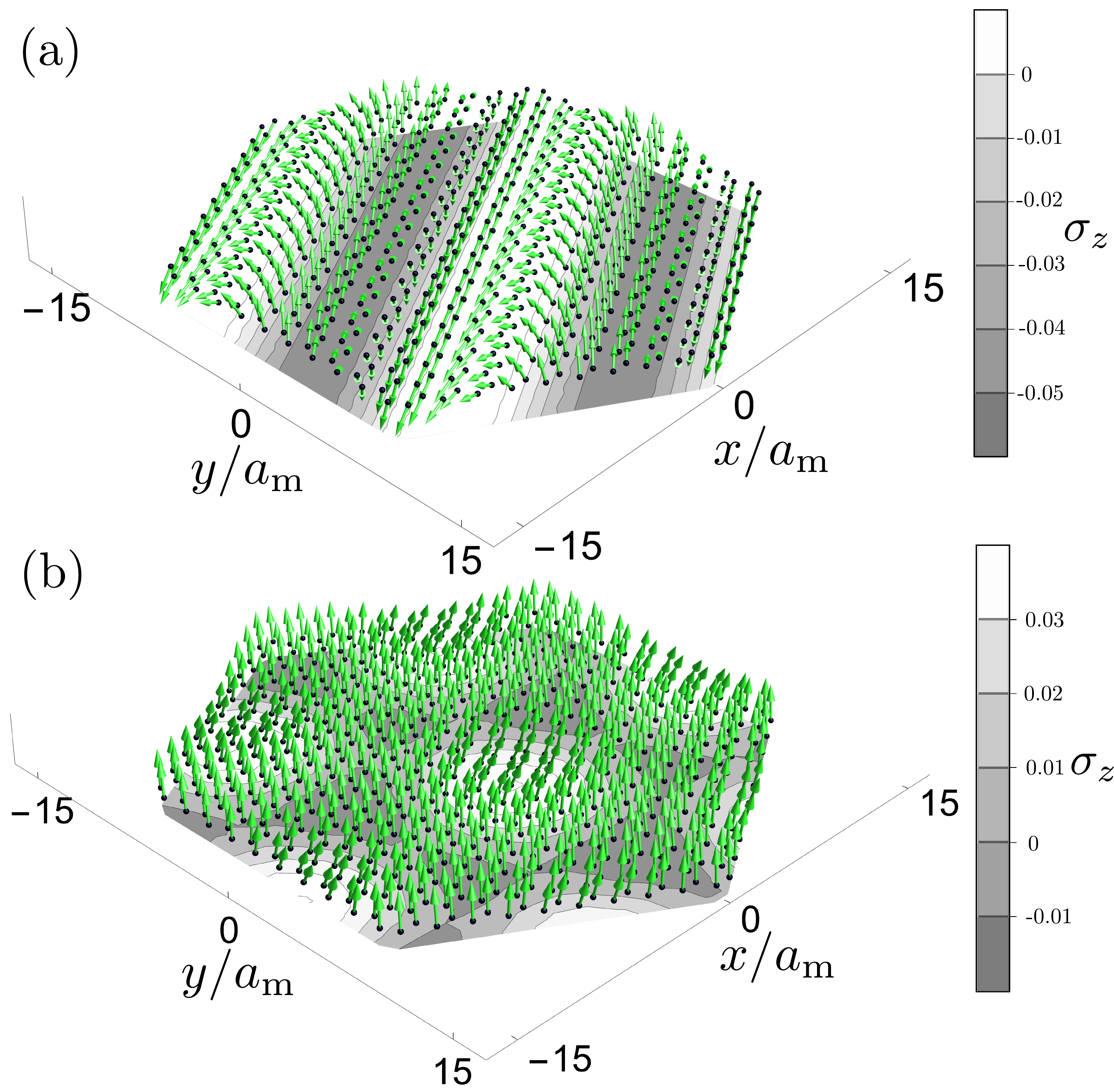}
\caption{Examples of inhomogeneous metallic spin-ordered phases for $U/W = 0.5$ (a) and $U = 0.33$ (b) at $\nu = 1.0$. Contours in gray show the local $z$-component of the spin, i.e. $\sigma_z = \frac{1}{2}\sum_{\tau} (\langle n_{i\tau\uparrow}\rangle - \langle n_{i\tau\downarrow}\rangle)$. The structure in (a) breaks $C_3$-symmetry and a single ordering vector has condensed along all three directions of the Bloch sphere. The structure shown in (b) preserves $C_3$ and three symmetry-related ordering vectors have condensed in all three directions. \label{fig:4}}
\end{figure}

\subsection{Phase diagram: inhomogeneous phases}\label{Subsec:inhomo}

As indicated in Fig~\ref{fig:2}, inhomogeneous metallic phases exist at all finite integer filling factors in the regime of low to intermediate interaction strengths, as expected for weak-coupling approaches. In Fig.~\ref{fig:2} these phases are indicated  by either orange ($U_v(1)$ preserving) or purple color ($U_v(1)$ breaking) as the gapless and inhomogeneous properties are common to all. However, the flavor order degree of freedom as well as the specific ordering vector(s) are found to depend on the particular filling factor and interaction strength. The variety of phases identified in the low to intermediate interaction regime implies a plethora of (near-)degenerate states in this region. This combined with the impact of finite size effects in self-consistent computations of inhomogeneous ground states causes significant convergence difficulties and the notion of simulations trapped in local minima in the extensive phase space cannot be dismissed. Nonetheless, the breaking of translation symmetry and the gapless energy spectrum are highly robust features across all nonzero integer fillings at these interaction strengths. 

Examples of typical inhomogeneous phases are shown in Fig.~\ref{fig:4}. Figure~\ref{fig:4}(a) displays the spin structure of the converged result at $U/W = 0.5$ and $\nu = 1$, where a single ordering vector has condensed in all three directions of the Bloch sphere. The spin modulations in the three directions have relative phase shifts resulting in a spin spiral with a net relative magnetization of $\langle S^z_i \rangle/\langle |\mathbf{S}_i| \rangle = 0.24$ per site. For lower interactions $U/W = 0.33$ at $\nu = 1$, the ground state exhibits the spin pattern shown in Fig.~\ref{fig:4}(b) with a net relative magnetization of $\langle S^z_i \rangle/\langle |\mathbf{S}_i| \rangle = 0.97$ per site accompanied by $C_3$ preserving ordering vectors for each spin direction.

\begin{figure}[bt]
\centering
\includegraphics[angle=0,width=1.0\linewidth]{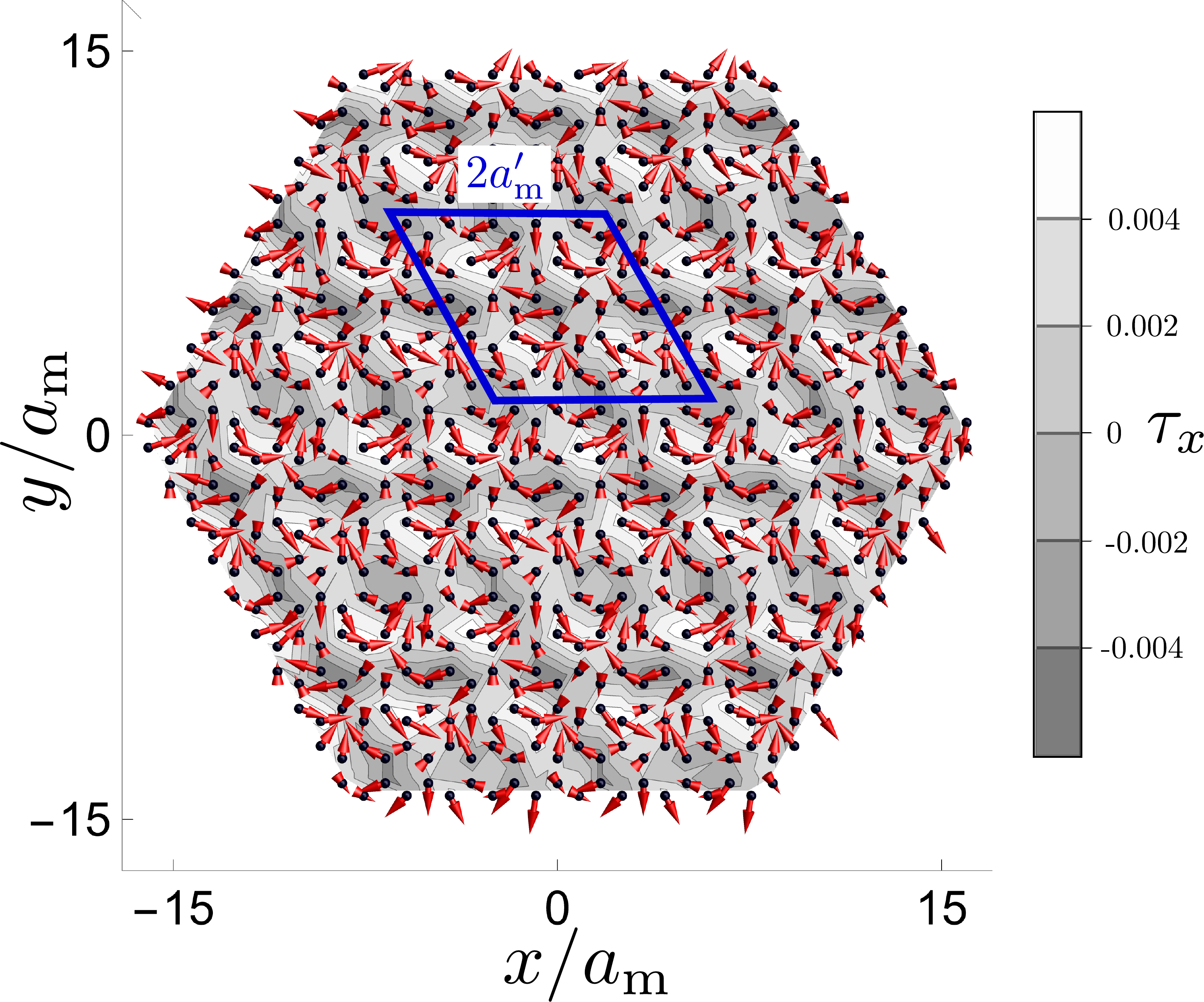}
\caption{Example of inhomogeneous metallic valley-ordered phase for $U/W = 0.33$ at $\nu = 2.0$. Contours in gray show the local $x$-component of the valley order, i.e. $\tau_x = \sum_{\sigma}\mathrm{Re}(\langle \cd_{i,+,\sigma}c_{i,-,\sigma}\rangle + \langle \cd_{i,-,\sigma}c_{i,+,\sigma}\rangle)$. The order is restricted to the $xy$-plane of the valley Bloch sphere and the unit cell is enlarged to $a'_{\rm m} = 2.5a_{\rm m}$ as indicated by the blue rhombus. \label{fig:5}}
\end{figure}

Figure~\ref{fig:5} shows an example of a valley-ordered phase exhibited at weak coupling at $\nu = 2.0$. This phase is characterized by spatially modulating intervalley coherence breaking both translation and $U_v(1)$ symmetry. The direction of this modulating intervalley coherence is restricted to the $xy$-plane of the valley Bloch sphere, i.e. the occupation of the two valleys is identical. In this case, the order is defined by $C_3$-related vectors of length $|{\mathbf{q}}_i| = 0.4 |{\mathbf{G}}_{\mathrm m}|$, where ${\mathbf{G}}_{\mathrm m}$ denote the reciprocal lattice vectors of the moir{\'e} Brillouin zone (mBZ). Interestingly, a modulated intervalley coherence phase has been also recently explored within HF studies of the BM model where it is found to be the preferred ground state across all nonzero integer fillings in the presence of a small $C_{3}$ breaking heterostrain~\cite{Kwan_2021,Wagner2022}. There, access to the graphene scale has identified this order as an incommensurate Kekul{\'e} pattern.

Finally, we stress that while the inhomogeneous phases discussed above are driven by a single degree of freedom, the order is generally inherited in the remaining degrees of freedom. An example is the negligible, yet finite, spin order ($\langle |\mathbf{S}_i| \rangle = 0.002$ per site) in the valley-ordered phase depicted in Fig.~\ref{fig:5}.

\begin{figure}[t]
\centering
\includegraphics[angle=0,width=1.0\linewidth]{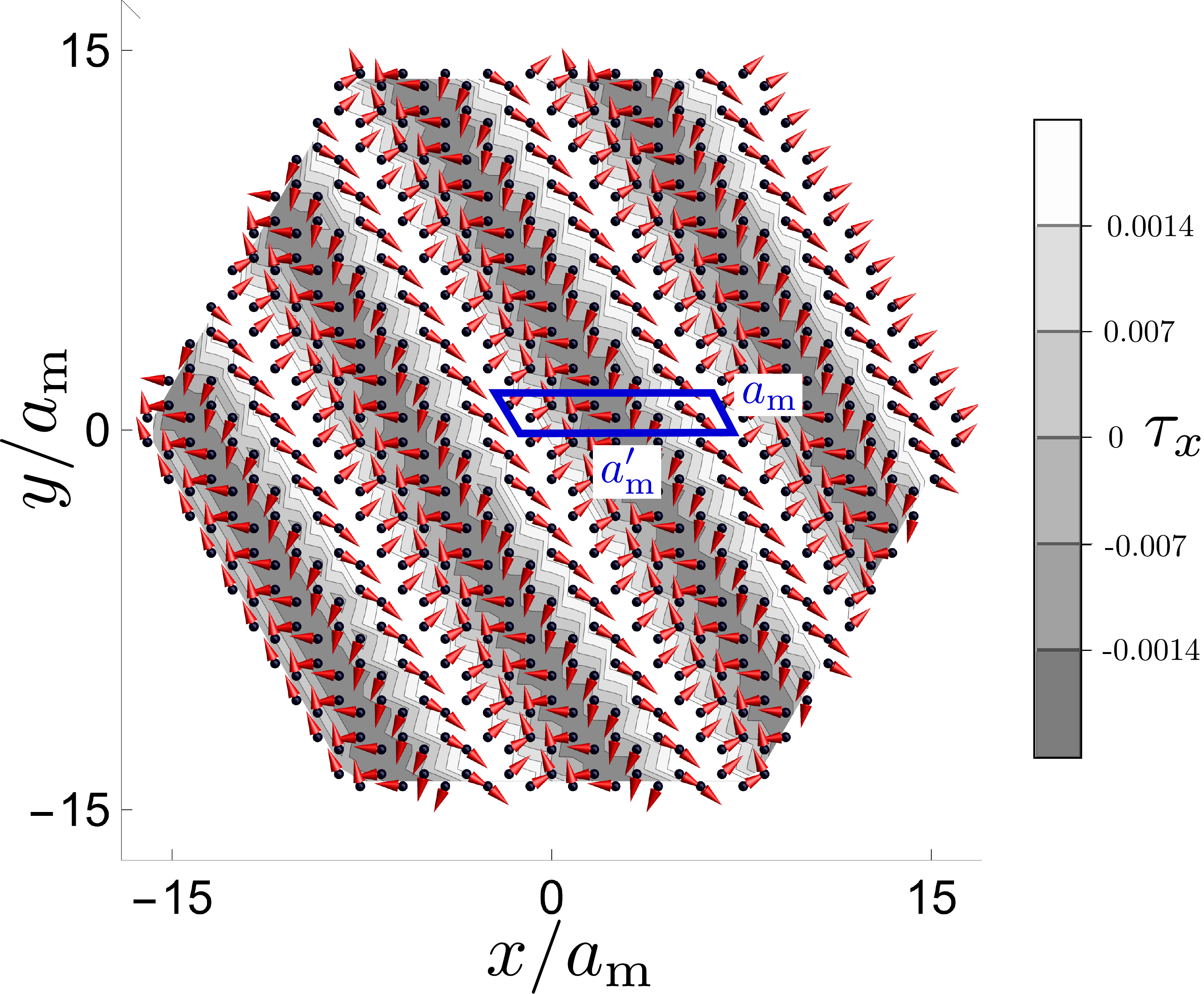}
\caption{Example of inhomogeneous metallic valley-ordered phase for $U/W = 0.33$ at $\nu = -0.5$. Contours in gray show the local $x$-component of the valley order, i.e. $\tau_x = \sum_{\sigma}\mathrm{Re}(\langle \cd_{i,+,\sigma}c_{i,-,\sigma}\rangle + \langle \cd_{i,-,\sigma}c_{i,+,\sigma}\rangle)$. The order is restricted to the $xy$-plane of the valley Bloch sphere defined by a single ordering vector of length $|\fq| = 0.2|\mathbf{G}_{\mathrm m}|$ ($a'_{\rm m} = 5 a_{\rm m}$) with a relative phase between the $x$- and $y$-direction yielding an intervalley coherent spiral.}\label{fig:6}
\end{figure}

\subsection{Modulated phases at half-integer filling}\label{Subsec:halfint}

Recently, Bhowmik {\it et al.}~\cite{Bhowmik2022} studied MATBG proximitized by a layer of WSe$_2$ and reported distinct features in magnetotransport and thermoelectricity consistent with ordered phases setting-in at half-integer filling factors, $\nu=\pm 0.5$ and $\nu=\pm 3.5$. These results were interpreted in terms of correlation-induced modulated spin- and charge-density wave order with doubled unit cell order at the moir{\'e} scale. We have computed the mean fields at $\nu=\pm 0.5$, and show the resulting intervalley coherent order at $\nu=-0.5$ and $U/W = 0.33$ in Fig.~\ref{fig:6}. As seen, the order is similar to the intervalley coherence spiral shown in Fig.~\ref{fig:5} with an additional breaking of $C_3$-symmetry. The intervalley order is restricted to the $xy$-plane of the valley Bloch sphere, and we find a well-defined ordering vector of length $|\fq| = 0.2 |\mathbf{G}_{\mathrm m}|$ with a relative phase between the $x$- and $y$-directions yielding an intervalley coherent spiral along $\fq$. Similar to the inhomogeneous phases discussed in the previous section, the phase shown in Fig.~\ref{fig:6} is metallic.

For larger interaction strengths at half-integer filling $\nu = \pm0.5$ the ground states remain periodically modulated metallic phases dominated by order in the valley degree of freedom. For example, at $U/W = 1.0$ and $\nu = -0.5$ the order is a modulated valley polarization preserving $U_v(1)$ symmetry, while intervalley coherence is present in the valley-ordered state at $\nu = +0.5$. However, it should be noted that, at these higher interaction strengths, significant order is also present in the spin degree of freedom with magnitudes smaller than but comparable to the valley order. Thus, the appearance of modulated ordered phases is a robust feature also at half-integer fillings of the current model, and this may be related to the unusual magnetotransport features detected in recent experiments~\cite{Bhowmik2022}.
 
\section{Discussion and Conclusions}

We have performed an unrestricted self-consistent HF study of a moir{\'e} lattice model designed for MATBG with both local and non-local interactions truncated to NN sites, obtained from Wannier projected Coulomb repulsion, and mapped out the preferred ordered phases as a function of interaction strength and the moir{\'e} flat-band filling factor. This model proposed by Kang and Vafek has been previously addressed e.g. from a  strong-coupling perspective and also studied via numerical methods including QMC simulations and density matrix renormalization group (DMRG) calculations~\cite{KangVafek2019,ClaraPRX2021,LiaoPRL2019,ChenMeng2021}. For example, Ref.~\cite{ChenMeng2021} recently studied the model in the flat-band, flavor-polarized limit at half-filling using a DMRG approach (corresponding to half filling of a single of the four decoupled band sectors discussed in Section~\ref{sec:QVHbands} with $H_0 = 0$). The study reports a first order phase transition from a stripe charge density-wave phase to the QVH phase at a critical assisted-hopping interaction of  $\alpha_c/U \approx 0.12$. Interestingly, the authors find a remarkably low von Neumann entropy of the QVH phase and argue therefore that a similar ground state should be well captured by a self-consistent mean-field description, a conjecture consistent with the results presented here even in the non-(assumed)-flavor-polarized, non-flat-band limit.

Furthermore, previous HF studies of the Kang-Vafek model restricted to charge neutrality, but including a NN tight-binding hopping term to the Hamiltonian, obtained the QVH phase and was shown to agree well with QMC simulations~\cite{ClaraPRX2021}. Here, we have extended such HF studies to include other electron filling factors and explored the resulting preferred ordered phases. As seen from Fig.~\ref{fig:2}, for interaction strengths large enough we obtain ferromagnetic spin/valley polarized  order coexisting with the QVH phase. This in turn gives rise to a particular filling-dependence of the moir{\'e} bands with Chern-insulator phases at integer filling factors, and presence of quantum anomalous Hall effect at odd filling factors. Thus far, experimental efforts have revealed a multitude of Chern-insulator phases in MATBG under the application of an external magnetic field \cite{Nuckolls2020, Wu2021, Pierce2021, Das2021}, however, it has proved challenging to observe topological signatures of the correlated insulators in the absence of external fields, most likely due to sample imperfections \cite{Grover2022, Stepanov2021}. Nevertheless, two observations of (nearly) quantized anomalous Hall effects at $\nu = +1,+3$ alongside a detailed inverse compressibility study down to zero field at $\nu = +1,+2,+3$ have been published recently \cite{Pierce2021, Serlin2020, Stepanov2021}. Interestingly, while many reports on field-stabilized Chern insulators in MATBG find sequences of $\{C,\nu\} = \{\pm4,0\},\{\pm3,\pm1\},\{\pm2,\pm2\},\{\pm1,\pm3\}$, the zero-field Chern insulators at $\nu = +1,+3$ ($\nu = +2$) find quantizations of $|C| = 1$ ($C = 0$) in agreement with the results presented here.

At weak to intermediate interaction strengths $U/W \lesssim 1$, the preferred ground states become inhomogeneous with density-wave order in spin/valley degrees of freedom. Similar inhomogeneous phases are found at certain half-integer filling factors, which appears consistent with recent experimental reports of ordered spatially-modulated phases at certain half-integer filling factors~\cite{Bhowmik2022}. We stress that the modulated phases discussed here are intrinsic, i.e. generated solely by the interactions. In actual samples external inhomogeneities from e.g. strain or twist-angle variations will lead to additional spatial variations of pinned order, possibly producing a mosaic of regions featuring different topological properties~\cite{Grover2022}. Inclusion of such effects, and a more detailed comparison to experiments in terms of the derived spectral properties and transport coefficients is beyond the scope of the current work. Here, we have focused on mapping out the filling-dependence of the preferred ordered phases of the present moir{\'e} lattice model. This already exhibits a plethora of different phases, some of which constitute fascinating examples of interaction-driven nontrivial topological states of matter.

\begin{acknowledgments}
We acknowledge useful discussions with E. Berg, M. H. Christensen, R. M. Fernandes, M. Geier, J. Kang, Z.-Y. Meng, and G. Wagner. C.N.B. acknowledges support by the Danish National Committee for Research Infrastructure (NUFI) through the ESS-Lighthouse Q-MAT.
\end{acknowledgments}
\bibliography{litlist}
\end{document}